\documentclass[12pt]{article} 
\pdfoutput=1

\usepackage[sectionbib]{natbib}
\usepackage[sectionbib]{chapterbib}

\usepackage{array,epsfig,fancyheadings,rotating}
\usepackage[]{hyperref}  
\usepackage{sectsty, secdot}

\usepackage{JASA_manu}

\usepackage{amsmath}
\usepackage{amssymb}
\usepackage{amsfonts}
\usepackage{multirow}
\usepackage{amsthm}
\usepackage{cleveref}
\usepackage{framed}
\usepackage{algorithm}
\usepackage{algorithmic}
\usepackage{subfigure}
\usepackage{epsfig}

\def\real{\mathbb{R}}
\def\cov{\mathrm{cov}}
\def\transpose{\mathrm{T}}
\def\tdomain{\mathcal{T}}
\def\bu{\boldsymbol{u}}
\def\bx{\boldsymbol{x}}
\def\bX{\boldsymbol{X}}
\def\hilbert{\mathbb{H}}
\def\bu{\mathbf{u}}
\global\long\def\bpsi{\boldsymbol{\psi}}

\global\long\def\btheta{\boldsymbol{\theta}}

\newtheorem{theorem}{Theorem}
\newtheorem{lemma}{Lemma}

\newtheorem{proposition}{Proposition}
\theoremstyle{definition}

\newtheorem{remark}{Remark}

\newtheorem{condition}{\it{Condition}}

\newcommand{\single}{\renewcommand{\baselinestretch}{1.2}\normalsize}
\newcommand{\double}{\renewcommand{\baselinestretch}{1.63}\normalsize}

\allowdisplaybreaks

\begin{document}

\title{Sparse Functional Principal Component Analysis in High Dimensions}

\author{Xiaoyu Hu and Fang Yao
	\\
	{\noindent\em\small School of Mathematical Sciences, Center for Statistical Science, Peking University, China}    }
\date{}
\maketitle

\single
\begin{center}
	\textbf{Abstract}
\end{center}
Functional principal component analysis (FPCA) is a fundamental tool and has attracted increasing attention in recent decades, while existing methods are restricted to data with {a single or finite number of} random functions (much smaller than the sample size $n$).
In this work, we focus on high-dimensional functional processes where the number of random functions $p$ is comparable to, or even much larger than $n$.
Such data are ubiquitous in various fields such as neuroimaging analysis, and cannot be properly modeled by existing methods. We propose a new algorithm, called sparse FPCA, which is able to model principal eigenfunctions effectively under sensible sparsity regimes.
While sparsity assumptions are standard in multivariate statistics, they have not been investigated in the complex context where not only is $p$ large, but also each variable itself is an intrinsically infinite-dimensional process. 
The sparsity structure motivates a thresholding rule that is easy to compute without nonparametric smoothing by exploiting the relationship between univariate orthonormal basis expansions and multivariate Kahunen-Lo\`eve (K-L) representations.
{We investigate the theoretical properties of the resulting estimators, and illustrate the performance with simulated and real data examples.}

\vspace*{.1in}


\pagenumbering{arabic} 

\double

\section{Introduction}\label{sec:intro}
Functional data have been  commonly encountered in modern statistics, and dimension reduction plays a key role due to the infinite dimensionality of such data.
As an important tool for dimension reduction, FPCA is optimal in the sense that the integrated mean squared error is efficiently minimized, which has wide applications in functional regression, classification and clustering \citep{rice1991estimating, yao2005functionala,yao2005functionalb,muller2005generalized,hall2006properties,hall2007methodology,horvath2012inference,dai2017optimal,wong2019partially}. 
Despite progress being made in this field, existing methods often involve a single or finite number of random functions. In this paper, we focus on modeling principal eigenfunctions of $p$ random functions where $p$ is comparable to, or even much larger than the sample size $n$, i.e., the number of subjects. Such data, which are referred to as the high-dimensional functional data, are becoming increasingly available in various fields, and examples can be found in neuroimaging analysis where various brain regions of interest (ROIs) are scanned over time for individuals. 

A typical example is the electroencephalography (EEG) data, see Section \ref{sec:realdata} for a description of the dataset that consists of $n=122$ subjects with 77 in the alcoholic group and 45 in the control group. For each subject, $p=64$ electrodes are recorded at $m=256$ time points for one-second interval, and classification using brain signals is often of interest.
In brain computer interface (BCI) applications, a widely adopted approach is to use spatial covariance matrices (averaged over time) as EEG signal descriptors and implement classification under Riemannian manifold perspective \citep{barachant2011multiclass,nguyen2018eeg,sabbagh2019manifold}.
However, due to the dynamic and non-stationary \citep{sun2014review} features of EEG signals, averaging over time may lead to lack of interpretation and/or loss of information in the original high-dimensional space, which is evidenced by results in Table 3 in Section \ref{sec:realdata}. Hence we aim to model the data directly and provide an efficient yet effective means to extract features from the original signals \citep{qiao2019functional,qiao2020doubly,solea2020copula}.

To deal with such high-dimensional processes, a straightforward way is to extract features through $p$ individual FPCAs and apply high-dimensional techniques to reduced variables. Nevertheless, this strategy has some drawbacks. First, it is computationally expensive since $p$ univariate FPCAs and additional high-dimensional methods are required. Second, one of the advantages of FPCA is to provide a parsimonious representation of data, while separate decompositions fail to model the correlation between processes and make the interpretation difficult. Moreover, the correlation among FPC scores from different processes may lead to multicollinearity in subsequent regression analysis (e.g., \citealp{muller2008functional}). Finally, there is no theoretical guarantee for collectively treating high-dimensional functional features, nor for the performance of subsequent analysis. Therefore, classical methods and results are no longer applicable, which motivates the study of scalable FPCA in high dimensions.

Recently there has been elevated interest in studying multivariate FPCA. 
A standard approach is to concatenate the multiple functions to perform univariate FPCA \citep[Chapter 8.5]{ramsay2005}. \citet{berrendero2011principal} performed a classical multivariate PCA for each value of the domain on which the functions are observed. 
\citet{chiou2014multivariate} proposed a normalized version of multivariate FPCA.
\citet{jacques2014model} introduced a method based on basis expansions, and \citet{happ2018multivariate} extended it to handle multivariate functional data observed on different (dimensional) domains. In the aforementioned works, the number of functional variables $p$ is considered finite and much smaller than the sample size $n$. As a consequence, these methods fail to deal with functional data in high dimensions  due to both computational and theoretical issues.

Likewise in multivariate statistics, the sample eigenvectors are inconsistent in high dimensions \citep{johnstone2009consistency}.
A typical strategy is to impose the sparsity assumption on eigenvectors or principal subspace \citep[among others]{zou2006sparse,shen2008sparse,vu2013minimax}. In particular, \citet{johnstone2009consistency} proposed an estimator based on diagonal thresholding that screens out variables with small sample variances. In spite of extensive literature for sparse PCA, the extension to high-dimensional functional processes is still  challenging, as the functional data are usually observed at grids with noise and the large $p$ leads to error accumulation. Moreover, there is no available notion of sparsity in the context of high-dimensional functional data where not only is $p$ large, but also each variable is an intrinsically infinite-dimensional process. 

{Our goal is to establish the parsimonious sparse FPCA which facilitates interpretation for high-dimensional functional data}. We begin with establishing the connection between the multivariate K-L expansion and univariate orthonormal basis representation for infinite-dimensional processes, which is a generalization of \citet{happ2018multivariate} assuming that each process has a finite-dimensional representation. 
The established relationship is flexible to allow any suitable basis expansions such as orthonormal B-spline basis and Wavelet basis.
Based on this relationship, our method avoids performing univariate FPCAs which are computationally expensive and introduce data-dependent uncertainty in high dimensions. 
The main contributions include coupling the sparsity concept in multivariate statistics with functional variables. While the sparsity is standard in multivariate statistics, there has been no attempt to generalize it to functional settings.
The sparsity structure motives us to adopt the thresholding technique which can identify important processes and avoid intensive computation. 
Moreover, we carefully investigate the theoretical properties of resulting estimators, as well as the complex interaction between the eigen problem and the sparsity regularization.
A phase transition phenomenon intrinsic to discretely observed functional data in terms of the sampling rate is revealed in this context. To our knowledge, this has not been discussed in literature and provides insight into consistent dimension reduction for discretely observed noisy functional data in high dimensions.

The remainder of the article is organized as follows. In \Cref{sec:sfpca}, we provide the sparsity assumption and introduce the proposed approach sparse FPCA (sFPCA). In \Cref{sec:thm}, we present the theoretical results for sFPCA under the sparsity regime. Simulation results for both trajectory recovery and classification are included in \Cref{sec:sim}, followed by an application to the EEG data in \Cref{sec:realdata}. More theoretical results, technical proofs and simulations are deferred to the Supplementary Material.

\section{Sparse FPCA in high dimensions}\label{sec:sfpca}

\subsection{Multivariate Karhunen-Lo\`eve expansion}

{We first give some notations used in the sequel. The boldface letters are used for denoting vectors, while the uppercase $\bX$ for the model and lowercase $\bx$ for the observed sample. For a vector $\bx=(x_1, \dots, x_p)^{\transpose}$, let $x_{(j)}$ denote the $j$th coordinate that is  non-increasingly ordered, such that $x_{(1)} \ge x_{(2)} \ge \dots \ge x_{(p)}$. For $n \in \mathbb{N}$ and two sequences of real numbers, $\alpha_n$ and $\beta_n$, $\alpha_n \approx \beta_n$ stands for $\alpha_n / \beta_n \to 1$, $\alpha_n \ll \beta_n$ for $\alpha_n/\beta_n \to 0$, $\alpha_n \gg \beta_n$ for $\alpha_n / \beta_n \to \infty$ and $\alpha_n \propto \beta_n$ denotes $0 < \alpha_n /\beta_n < \infty$ as $n \to \infty$.
}

Suppose that the functional data are $\bX(t) = (X_1(t), \dots, X_p(t))^{\transpose}$ and each $X_j(\cdot) \in L^2(\tdomain)$ is a square-integrable random function defined on a compact interval $\tdomain = [0, 1]$ with continuous mean and covariance functions. Let $\hilbert$ denote a Hilbert space of $p$-dimensional vectors of functions in $L^2(\tdomain)$, equipped with the inner product $<\boldsymbol{f}, \boldsymbol{g}>_{\hilbert} = \sum_{j=1}^p \int_{\tdomain} f_j(t)g_j(t) dt$ and the norm $\|\cdot\|_{\hilbert}=<\cdot,\cdot>_{\hilbert}^{1/2}$. Without loss of generality (w.l.o.g.), we assume that all processes are centered, i.e., $E\{X_j(t)\} = 0$. Define the covariance function $G(s, t) = E\{\bX(s) \bX(t)^{\transpose}\} = \{G_{jk}(s, t)\} \in \real^{p\times p} $.

According to the multivariate Mercer's theorem \citep{balakrishnan1960,kelly1960}, there exists a complete orthonormal basis $\{\bpsi_k(t): k\geq 1\}$ and the corresponding sequence of eigenvalues $\{\lambda_k > 0 : k\geq 1\}$ such that $G(s, t)$ has the representation $G(s, t) = \sum_{k=1}^{\infty} \lambda_k \bpsi_k(s) \bpsi_k(t)^{\transpose}$, where $< \bpsi_{k_1}(t),\bpsi_{k_2}(t)>_{\hilbert} = \delta_{k_1 k_2}$, where $\delta_{k_1 k_2}$ is $1$ if $k_1 = k_2$ and $0$ otherwise, and $\lambda_1 \ge \lambda_2 \ge \cdots \ge 0$. 
Accordingly, the multivariate K-L expansion is
$\bX(t) = \sum_{k=1}^{\infty} \eta_{k} \bpsi_k(t)$,
where $\bpsi_k(t) = (\psi_{k1}(t), \dots, \psi_{kp}(t))^{\transpose}$ and the scores $\eta_{k} = < \bX, \bpsi_k >_{\hilbert}$ are random variables with mean zero and variances $E(\eta_{k}^2) = \lambda_k$. It leads to a single set of scores for each subject, which serves as a proxy of multivariate functional data. In contrast, the univariate Karhunen-Lo\`eve expansion is
$X_{j}(t) = \sum_{k=1}^{\infty}\xi_{jk}\phi_{jk}(t)$,
where $\xi_{jk}=\int_{\tdomain} X_{j}(t) \phi_{jk}(t)dt$ and $\phi_{jk}(t)$ are eigenfunctions satisfying $\int_{\tdomain} \phi_{jk_1}(t)\phi_{jk_2}(t)dt=\delta_{k_1k_2}$. To avoid the ambiguity, we refer to $\bpsi_{k}(t)$ and $\phi_{jk}(t)$ as multivariate and univariate eigenfunctions, respectively. Clearly the main difference between these two expansions is that the $\bpsi_k(t)$ are vector-valued while the scores $\eta_{k}$ are scalars, which allows a parsimonious representation of data and the same structure for each subject. Our focus of interest is to establish consistent estimators for $\bpsi_k(t)$, and as a consequence, the scores $\eta_{k}$ and parsimonious data recovery are obtained. 

\subsection{Basis representation for Karhunen-Lo\`eve expansion}

In high dimensions, computational tractability is one of practical considerations.
Either pre-smoothing \citep{ramsay2005} or post-smoothing \citep{yao2005functionala} method for FPCA is computationally prohibitive when $p$ is large {\citep{xue2021functional}, which is further discussed in Remark \ref{rem:complexity}}. A remedy is to represent functional processes via a set of orthonormal basis, consequently, the covariance/eigenfunctions are expressed and estimated accordingly \citep{rice2001nonparametric,james2000principal}. We derive the relationship between univariate basis expansions and multivariate K-L representations in \Cref{prop1} for intrinsically infinite-dimensional processes, setting stage for the proposed methodology.
\begin{proposition}\label{prop1}
	Assume that $\bX \in \hilbert$. {Given a complete and orthonormal basis $\{b_l(t), l \ge 1\}$ in $L^2(\tdomain)$, the representation for each random process is $X_j(t)= \sum_{l=1}^{\infty} \theta_{jl} b_{l}(t)$, where $\theta_{jl} = \int_{\tdomain} X_j(t)b_l(t)dt $ and the sum converges in the mean square sense. Let $\bpsi_k$ and $\lambda_k$ be eigenfunctions and corresponding eigenvalues of the covariance operator of $\bX$. By Parseval's identity, denote $\psi_{kj}(t) = \sum_{l=1}^{\infty} u_{kjl} b_{l}(t)$ where $u_{kjl} = \int_{\tdomain} b_{l}(t)\psi_{kj}(t)dt$. We have
	\begin{equation}\label{eq:relation}
	\sum_{j^{\prime}=1}^{p}\sum_{l^{\prime}=1}^{\infty} \cov(\theta_{jl}, \theta_{j^{\prime}l^{\prime}})u_{kj^{\prime}l^{\prime}} = \lambda_k u_{kjl}, j = 1, \dots, p, k, l = 1, 2, \dots,
	\end{equation}
	with the sum converging absolutely and the scores $\eta_k = \sum_{j=1}^{p} \sum_{l=1}^{\infty} u_{kjl} \theta_{jl}$ with the sum converging in the mean square sense.} 
\end{proposition}

By contrast, \citet{happ2018multivariate} gave a similar relationship under the assumption of finite-dimensional representations. \Cref{prop1} is a generalization in line with the intrinsically infinite-dimensional nature of functional data.
Accordingly, the $j$th component of eigenfunctions $\bpsi_k$ can be expressed as a linear combination of bases $\{b_l: l\geq 1\}$ with generalized Fourier coefficients $\{u_{kjl}: l\geq 1\}$ obtained from \eqref{eq:relation} and that the scores $\eta_k$ are linear combinations of basis coefficients $\{\theta_{jl}: j=1,\dots,p;l=1,\dots,\infty\}$.

\Cref{prop1} allows arbitrary basis expansions incorporating a set of {pre-specified} basis (e.g., {orthonormal} B-splines, wavelets) or data-driven basis (i.e., eigenfunctions). Although eigenfunctions can be estimated from data, it is inadvisable to employ univariate FPCA which is computationally prohibitive for large $p$ and introduce data-dependent uncertainty. Therefore, we adopt {pre-specified} basis functions to represent the trajectories and covariance/eigenfunctions \citep{rice2001nonparametric,james2000principal}. W.l.o.g., we use a common complete and orthonormal basis $\{b_l: l \geq 1\}$ in $L^2(\tdomain)$ for $p$ processes and do not pursue other complicated basis-seeking procedures that are peripheral to the key proposal. Let the underlying random functions be expressed as $X_{j}(t) = \sum_{l=1}^{\infty} \theta_{jl} b_l(t)$, where the coefficients $\theta_{jl} = \int_{\tdomain} X_{j}(t)b_l(t) dt$ are random variables with mean zero and variances $E(\theta_{jl}^2)=\sigma_{jl}^2$, and we refer to the total variability of the $j$th process as its energy denoted by $V_j = \sum_{l=1}^{\infty} \sigma_{jl}^2 < \infty $. It is necessary to regularize infinite-dimensional processes, and a natural means is truncation that serves as a sieve-type approximation. The size of truncation may diverge with the sample size $n$, which maintains the nonparametric nature of the proposed method. Denote the number of basis functions by $s_{nj}$, also referred to as the truncation parameter of the $j$th process when no confusion arises, $j=1, \dots, p$. It suffices to use a common $s_n$ for the method development and theoretical analysis, assuming $s_{nj} \asymp s_n$. 

Through \Cref{prop1}, the multivariate FPCA can be transformed into performing the classical PCA on the covariance matrix of all basis coefficients. 
Moreover, this motivates an easy-to-implement estimation procedure under sensible sparsity regimes described in \Cref{subsec:sparsity}.

\begin{remark}\label{rem:basis}
{Pre-specified} basis expansion is a fairly popular method to deal with functional data, see \citet{james2000principal}, \citet{ramsay2005} and \citet{koudstaal2018multiple}, among others.
Although the \Cref{prop1} is presented using the same set of orthonormal basis functions for $p$ random processes to simplify the exposition, our method is applicable to the general case of different bases (not necessarily orthonormal) and/or domains. 
{Such generality results ensure that the random processes could lie in different Hilbert spaces, and we shall need to choose suitable bases to represent each process.
For non-orthonormal bases, the estimation algorithm can still be applied by taking the inner product matrix into consideration.} 
\end{remark}

\subsection{Sparsity regimes}\label{subsec:sparsity}

To our knowledge, there is no available notion of sparsity in the context of FPCA for high-dimensional cases where $p$ is large, though the sparsity of principal eigenvectors or subspace \citep{vu2013minimax} in multivariate statistics is well defined. The formulation of sparsity in our problem is nontrivial. First, FPCA depends on vector-valued eigenfunctions, not vectors anymore. Second, functional data are usually discretely observed with  error, which leads to more challenging estimation and data recovery due to error accumulation in high dimensions. Therefore, we aim to reduce the dimensionality from $p$ to a much smaller one. To succeed, the total energy of data should be concentrated in a smaller number of processes. To achieve this, we need additional structures for high-dimensional functional data.

For the moment, we first review a typical decay assumption for univariate functional data \citep{koudstaal2018multiple}. Recall that $\sigma_{jl}^2 = E(\theta_{jl}^2)$ where $\theta_{jl} = \int_{\tdomain} X_j(t)b_l(t)dt$ is the basis coefficient of $X_{j}$. Assume for adequately large $s_n$,
\begin{eqnarray}\label{decay}
\sigma_{j(l)}^2 & = & O\{l^{-(1+2\alpha)}\},~~ l \leq s_n, \nonumber\\
\sigma_{jl}^2 & = & O\{l^{-(1+2\alpha)}\},~~ l>s_n,
\end{eqnarray}
uniformly in $j=1,\dots,p$, where $\alpha>0$ and $\sigma_{j(l)}^2$ denote the ordered values such that $\sigma_{j(1)}^2 \geq \sigma_{j(2)}^2 \geq \cdots$. This assumption ensures that the bulk of signals in each process are contained in the largest $s_n$ coordinates, {while the location and the order of these coordinates are unknown {\em a priori} for spatial adaptation \citep{donoho1994ideal}.
Such relaxation is more realistic for pre-specified basis and suitable for modeling functions with striking local features, see Figure 1 in \citet{koudstaal2018multiple} for graphical illustration. }


The decay condition \eqref{decay} is not enough to handle high-dimensional settings since it does not provide any regularization for the high dimensionality $p$. 
Recall that $\mathbf{V}=(V_1, \dots, V_p)^{\transpose}$ and $V_j = \sum_{l=1}^{\infty} \sigma_{jl}^2$ is the total energy of the $j$th process.
In the following, the sparsity is assumed for the high-dimensional vector $\mathbf{V}$, which is shown to be reasonable in practice as illustrated in \Cref{sec:realdata}.  


\emph{Weak $l_q$ sparsity}. A typical situation of interest is to incorporate processes with small energies that decay in a nonparametric manner. To be specific, assume that for some positive constant $C>0$,
\begin{equation}\label{sparsity}
V_{(j)} \leq Cj^{-2/q}, ~~j=1, \dots, p,
\end{equation}
where $0<q<2$ determines the sparsity level, i.e., smaller $q$ entails sparser processes. Consequently, the total energy is concentrated in the leading processes with large energies. Thus, a reasonable assumption is
\begin{equation}\label{bdecay}
\begin{split}
\sigma_{(j)(l)}^2 = O\{j^{-2/q}l^{-(1+2\alpha)}\},~~ l \leq s_n, \\
\sigma_{(j)l}^2 = O\{j^{-2/q}l^{-(1+2\alpha)}\},~~ l>s_n,
\end{split}
\end{equation}
where $\sigma_{(j)(l)}^2$ is the $l$th largest variance of coefficients for the process with energy $V_{(j)}$, and the extra term $j^{-2/q}$ in comparison with \eqref{decay} is due to the sparsity assumed in \eqref{sparsity}.

To summarize, different from the multivariate case, functional weak $l_q$ sparsity contain two types of decay: within processes determined by $\alpha$, and between processes determined by $q$. 
The decay within processes means that the variances of coefficients exhibit certain sparsity, while the decay between processes depicts the sparsity assumption on the high-dimensional energy vector $\mathbf{V}$.
The within-process sparsity is standard for univariate functional data \citep{koudstaal2018multiple}, while the between-process sparsity is for the first time specified to 
regularize the high dimensionality $p$ in the context of functional data. 
{Note that another type of sparsity, the $l_0$ sparsity in the sense of $\|\mathbf{V}\|_0 = g \ll p$, is also assumed for completeness which is investigated and available in the Supplementary Material for space economy.}

\subsection{Proposed thresholding estimation and recovery} \label{subsec:estimation}
Distinguishing from existing works, we aim to model eigenfunctions of $p$ random processes where $p \gg n$. {The standard FPCA methods, such as \citet{happ2018multivariate}, are no longer applicable due to computational and theoretical issues in high dimensions}, as illustrated in \Cref{sec:sim,sec:realdata}. In this section, we propose a unified framework to perform sparse FPCA based on the relationship declared in \Cref{prop1}.

Let $\{\bx_i(t): i=1,\dots,n \}$ be independent and identically distributed (i.i.d.) realizations from $\bX(t)$, where $\bx_i(t) = (x_{i1}(t), \dots, x_{ip}(t))^{\transpose}$. In reality, we do not observe the entire trajectories $x_{ij}$ but some noisy measurements,
$y_{ijk} = x_{ij}(t_{k}) + \epsilon_{ijk}, t_{k} \in \tdomain$,
where $\epsilon_{ijk}$ is measurement error independent of $x_{ij}$ with mean zero and variance $\sigma^2$, $i=1, \dots, n, j=1, \dots, p, k = 1, \dots, m$. For the sake of simplifying statements, we assume that the grid is regular, i.e., $t_k = k/m$, while our methodology can be directly applied to more general grid structures. The extremely sparse case when only a few measurements are available for each trajectory \citep{yao2005functionala} is beyond the scope of this article and can be investigated for future study.

According to the \Cref{prop1}, we first perform basis expansions for all processes based on discrete observations. Let $I_k=((k-1)/m, k/m], k=2, \dots, m$ and $I_1 = [0, 1/m]$, we define $y_{ij}^{*}(t)=y_{ijk}$ for $t \in I_k$, and define $x_{ij}^{*}$, $\epsilon_{ij}^{*}$ similarly. Observe that $y_{ij}^{*}(t) = x_{ij}^{*}(t) + \epsilon_{ij}^{*}(t)$, and projecting $y_{ij}^{*}(t)$ onto the orthonormal basis $b_l(t)$ yields
$
\hat{\theta}_{ijl} = \tilde{\theta}_{ijl} + \tilde{\epsilon}_{ijl}, ~~ l=1, \dots, s_n,
$
for a suitable choice of $s_n$, where {$\hat{\theta}_{ijl} = \sum_{k=1}^m y_{ijk}b_l(t_k)/m$} are estimated basis coefficients and $\tilde{\epsilon}_{ijl}$ is independent of $\tilde{\theta}_{ijl}$ with mean zero and variance $\tilde{\sigma}^2 = E(\tilde{\epsilon}_{ijl}^2) =\sigma^2 m^{-1}+O(m^{-2})$ due to discretization. We emphasize that our method avoids intensive computation by using basis expansion and thresholding. The impact of noise/discretization on resulting estimators is theoretically analyzed in \Cref{sec:thm}. 

Assume that $\theta_{ijl}$ and $\epsilon_{ijk}$ are jointly Gaussian.
Therefore, we conclude that $\hat{\sigma}_{jl}^2 \sim (\sigma^2m^{-1} + \tilde{\sigma}_{jl}^2) \chi_{n}^2/n$ where $\hat{\sigma}_{jl}^2 = n^{-1}\sum_{i=1}^n \hat{\theta}_{ijl}^2$ and $\tilde{\sigma}_{jl}^2 = E(\tilde{\theta}_{ijl}^2)$.
For the method development, it suffices to use $\sigma^2/m$ as an approximation of $\tilde{\sigma}^2$ to construct our estimators.
The difference between $\tilde{\sigma}_{jl}^2$ and $\sigma_{jl}^2$ is negligible for large $m$, and large values of $\sigma_{jl}^2$ are prone to have large sample variances $\hat{\sigma}_{jl}^2$. The idea is to include only the variables with largest sample variances. Thus, we perform the coordinate selection as follows,
\begin{equation}\label{eq:threshold}
\hat{I} = \{ (j,l), j=1,\dots,p; l=1,\dots,s_n : \hat{\sigma}_{jl}^2 \geq m^{-1}\sigma^2(1 + \alpha_n) \},
\end{equation}
where $\alpha_n = \alpha_0 \{n^{-1} \log(ps_n)\}^{1/2}$, $\alpha_0>\sqrt{12}$ is a suitable positive constant for theoretical guarantees \citep{johnstone2009consistency}. The choice of $\alpha_n$ is based on the concentration result of basis coefficients, and the number of basis $s_n$ comes from the sieve-like truncation for functional processes. When $l>m^{1/(2\alpha+1)}$ or $j>m^{q/2}$ the signals decrease rapidly below the noise level. We expect that the proposed strategy retains only sizable signals and forces the rest to zero leading to the desired model parsimony.


Denote the retained coefficients by $\btheta_{\hat{I}} = (\theta_{jl},(j,l) \in \hat{I})^{\transpose}$. Let $S_{\hat{I}} = n^{-1} \sum_{i=1}^n \hat{\btheta}_{i\hat{I}} \hat{\btheta}_{i\hat{I}}^{\transpose}$ be the sample covariance matrix. Based on \Cref{prop1}, we  perform multivariate PCA on $S_{\hat{I}}$ to yield principal eigenvectors $\hat{\bu}_{k}, k = 1, \dots, r_n$. Finally, we transform the results back to functional spaces,
\begin{equation*}
\hat{\psi}_{kj}(t) = \sum_{l:(j,l)\in \hat{I}} \hat{u}_{kjl} b_l(t), ~~ \hat{\eta}_{ik} = \sum_{(j,l):(j,l)\in\hat{I}} \hat{u}_{kjl} \hat{\theta}_{ijl}, ~~\hat{\bx}_i^{r_{n}}(t) = \sum_{k=1}^{r_n} \hat{\eta}_{ik} \hat{\bpsi}_k(t).
\end{equation*}
for $j=1, \dots, p, k = 1, \dots, r_n$. Let $N_j$ be the number of retained coefficients for the $j$th process. 
Apparently, $N_j=0$ implies that elements of the $j$th block of $\hat{\btheta}$ satisfy $\hat{\theta}_{jl} \notin \hat{\btheta}_{\hat{I}}$ for all $l = 1, \dots, s_n$, then each element of the $j$th block of $\hat{\mathbf{u}}_k$ equals to zero, $\hat{\psi}_{kj}(t) \equiv 0, k =1,\dots,r_n$, i.e., the $j$th random process will be ruled out. Otherwise for $N_j >0$, there exists at least one element of $j$th block of $\hat{\btheta}$ satisfying $\hat{\theta}_{jl} \in \hat{\btheta}_{\hat{I}}$, then the $j$th random process will be retained. The implementation algorithm is summarized below.

\begin{algorithm}[t!]
	 \caption{The algorithm for sFPCA.}
		
	Generally, denote $\bar{y}_{j}(t) = n^{-1}\sum_{i=1}^n y_{ij}^{*}(t)$ and $\tilde{y}_{ij}(t) = y_{ij}^{*}(t) - \bar{y}_{j}(t)$.
	
	(i) \emph{Projection and truncation.} Project $\tilde{y}_{ij}(t)$ onto the orthonormal basis functions $b_l(t)$ to yield $ \hat{\theta}_{ijl} = \int_0^1 \tilde{y}_{ij}(t)b_l(t) dt, j=1, \dots, p, l=1, \dots, s_n$.
	
	(ii) \emph{Thresholding.} Calculate the sample variances $\hat{\sigma}_{jl}^2$ of $\hat{\theta}_{ijl}$ and perform the subset selection based on the rule, 
	\[	\hat{I} = \{ (j,l), j=1,\dots,p; l=1,\dots,s_n : \hat{\sigma}_{jl}^2 \geq m^{-1}\sigma^2(1 + \alpha_n) \},	\]
	where $\alpha_n = 4 \{n^{-1} \log(ps_n)\}^{1/2}$ in our numerical studies.
	
	(iii) \emph{Eigen-decomposition and transformation.} Calculate the sample covariance matrix $S_{\hat{I}}$ of retained coefficients $\hat{\theta}_{\hat{I}}$. Perform PCA on $S_{\hat{I}}$ to yield principal eigenvectors $\hat{\bu}_{k}$, $k=1,\dots, r_n$, then calculate 
	\[
	\hat{\psi}_{kj}(t) = \sum_{l:(j,l)\in \hat{I}} \hat{u}_{kjl} b_l(t), ~~ \hat{\eta}_{ik} = \sum_{(j,l):(j,l)\in\hat{I}} \hat{u}_{kjl} \hat{\theta}_{ijl}, ~~\hat{\bx}_i^{r_{n}}(t) = \bar{\mathbf{y}}(t) + \sum_{k=1}^{r_n} \hat{\eta}_{ik} \hat{\bpsi}_k(t),
	\]
	where $\bar{\mathbf{y}} = (\bar{y}_1, \dots, \bar{y}_p)^{\transpose}$.
	
\end{algorithm}

\begin{remark}\label{rem:tuning}
	In practice, the variance $m^{-1}\sigma^2$ is usually unknown, we replace it by a quantile estimator $Q_{\rho}(\hat{\sigma}_{jl}^2: j=1,\dots,p, l=1,\dots,s_n)$ as suggested by \citet{koudstaal2018multiple}, where $Q_{\rho}(\mathbf{z})$, $0<\rho <1$, is the $100\rho$th sample quantile of sorted values in a vector $\mathbf{z}$. We also propose an objective-driven method to choose the parameter $\rho$ which controls the desired sparsity level, the truncation $s_n$ and the number of principal components $r_n$. For unsupervised problems, $\rho$ may be determined by a trade-off between the quality of recovery and model complexity, i.e., the number of retained processes, while we use $K$-fold cross-validation to choose $s_n$ and the fraction of variance explained to choose $r_n$ for reduced computation. If one considers a supervised problem, such as regression or classification, parameters $\rho, s_n$ and $r_n$ may be tuned by $K$-fold cross-validation to minimize the prediction/classification error. From our theoretical analysis and numerical experience, as a practical guidance, one may choose an adequate $s_n$ to characterize the features and mainly focus on choices of $\rho$ and $r_n$. More details and empirical evidence are offered in \Cref{sec:sim}.
\end{remark}

\begin{remark}\label{rem:complexity}
	To illustrate the computational advantage of our algorithm, we examine the order of computation complexity for estimation of covariance and eigenstructure, in contrast to that of HG method \citep{happ2018multivariate} and $p$ univariate FPCAs. The HG method operates with $O(np^2s_n^2+p^3s_n^3)$ complexity, which scales poorly for high-dimensional functional data.  The univariate FPCA with either presmoothing \citep{ramsay2005} or post-smoothing \citep{yao2005functionala} requires computation of order $O(npm^2+ pm^3)$ that is  fairly intensive for densely observed high-dimensional functional data. Our method retains at most $N=\sum_{j=1}^p N_j$ non-zero coordinates, where $N \ll ps_n$ almost surely according to \Cref{lemma:card}. Thus, our procedure operates with the complexity at the order of $O(nps_n + nN^2 + N^3)$, which achieves considerable computational savings and is demonstrated in the numerical studies in \Cref{sec:sim,sec:realdata}.
\end{remark}

We stress that the analysis of functional data is more challenging than that of multivariate data in high dimensions. First, since functional data are recorded at a grid of points, the estimation error from observed discrete version to functional continuous version needs to be investigated with care. Second, most literature assumed the spiked covariance model for sparse PCA, while it is not valid for functional data that has potentially infinite rank. Third, as discussed in \Cref{subsec:sparsity}, the variances of coefficients involve two types of decay: within processes, i.e., $\alpha$, and between processes, i.e., $q$.

\section{Theoretical Properties}\label{sec:thm}

{
In this section, we focus on the consistency of eigenfunction estimates under the weak $l_q$ sparsity and more results for recovery of FPC scores and trajectories are deferred to the Supplementary Material for space economy.
To begin with, we state key conditions necessary for theoretical analysis, in which Conditions \ref{cd:gauss}-\ref{cd:m} concern properties of underlying processes and how the functional data are sampled/observed.   
}
\begin{condition}\label{cd:gauss}
	The basis coefficients $\theta_{ijl}$ and measurement errors $\epsilon_{ijk}$ are jointly Gaussian.
\end{condition}

\begin{condition}\label{cd:xsmooth}
	The sample paths are Lipschitz continuous, i.e., $|X_j(t)-X_j(s)|\leq L_{X_j}|t-s|$, and assume $E(L_{X_j}^2) < \infty$ for $j=1,\dots,p$. Moreover, $E(\theta_{jl}^4) \leq C\{E(\theta_{jl}^2)\}^2$.
\end{condition}

{The Gaussian assumption is needed to determine the constant $\alpha_0$ in the thresholding value $\alpha_n$ \citep{donoho1994ideal,koudstaal2018multiple}.} Conditions \ref{cd:gauss} and \ref{cd:xsmooth} imply that $X_j$ is a Gaussian process with continuous sample paths, while the moment conditions are standard in FDA literature \citep{hall2007methodology, kong2016partially}. Next condition prevents the spacing between adjacent eigenvalues from being too small and implies that $\lambda_k \geq Ck^{-a}$.

\begin{condition}\label{cd:egap}
	For $a>1$ and $C>0$, $\lambda_{k}- \lambda_{k+1} \geq Ck^{-a-1}, k \geq 1$.
\end{condition}

\begin{condition}\label{cd:sampling}
	Let $t_k = k/m$, and the $\{t_{k}, k=1, \dots, m\}$ are considered deterministic and ordered increasingly.
\end{condition}

\begin{condition}\label{cd:m}
	The sampling rate satisfies $m = O(n^{\gamma})$ for $\gamma > (1-\beta)/2$.
\end{condition}

To simplify the exposition, we assume that the data are equally spaced. The algorithm can be readily generalized to more general designs by defining $\delta = \sup_{i,j,k}\{t_{ij,k+1} - t_{ij,k}\}$ and $m = \inf_{i,j} m_{ij}$ and assuming $\delta=O(1/m)$.
Regarding the sampling frequency $m$, it should be large enough to control the discretization error such that $\tilde{\sigma}_{jl}^2/\sigma_{jl}^2 \to 1$.
Note that the Condition \ref{cd:m} is milder than that imposed by \citet{kong2016partially}. We shall see from later theorems that this assumption on sampling rate plays an indispensable role in approximation/estimation error. 
The number of functional processes $p$ is allowed to be ultrahigh.  

\begin{condition}\label{cd:p}
	$p = O\{\exp(n^{\beta})\}$ for $0< \beta <1$.
\end{condition}

It is standard to assume that $s_n$ should not be too small to capture the significant coordinates. Moreover, it should not be too large for reliable concentration results of sample variances of $\theta_{ijl}$, which provides theoretical foundation for establishing the thresholding rule.
Thus, it suffices to have an adequately large $s_n$ which is a useful guidance in practice. Moreover, we impose Lipschitz continuity on the basis functions without loss of generality.

\begin{condition}\label{cd:sn}
	The truncation number $ \left(m^{-1}\sqrt{\log p/n}\right)^{-1/(2\alpha+1)}\ll s_n = O(p)$.
\end{condition}

\begin{condition}\label{cd:blip}
	The basis functions are Lipschitz continuous, i.e., $|b_l(t)-b_l(s)|\leq L|t-s|$ for all $l=1, \ldots, s_n$. 
\end{condition}

We control the number of principal components $r_n$ such that it is not too large for increasingly unstable estimates. Conditions \ref{cd:lqrn1} and \ref{cd:lqrn2} concern the approximation error and estimation error, respectively.

\begin{condition}\label{cd:lqrn1}
	$r_n^{a+1} \max\left\{ g_n^{1/2-1/q+\delta}, \left(m^{-1}\sqrt{\log p/n}\right)^{\alpha/(2\alpha+1)} \right\} = o(1)$ for some $\delta>0$.
\end{condition}

\begin{condition}\label{cd:lqrn2}
	$\max\left\{ r_n^{a+1}n^{-1/2}, r_n^{a+1}g_n^{1/2} m^{-1} \right\} = o(1)$.
\end{condition}

In the asymptotic analysis, we consider the approximation error caused by truncation/thresholding as well as the statistical estimation error. For the eigenfunctions, one has the following decomposition:
$
\|\bpsi_k - \hat{\bpsi}_{k}\|_{\hilbert} \leq \|\bpsi_k - \tilde{\bpsi}_{k} \|_{\hilbert} + \|\tilde{\bpsi}_{k} -  \hat{\bpsi}_{k}\|_{\hilbert},
$
where $\tilde{\bpsi}_{k}$ are the eigenfunctions of thresholded processes $\tilde{\bX}$ with $\tilde{X}_j(t) = \sum_{l: (j,l) \in \hat{I}} \theta_{jl} b_l(t)$.
The first term on the right-hand side could be viewed as the approximation error, while the second term is interpreted as the estimation error. Recall that $N_j$ is the number of retained coefficients $\hat{\theta}_{ijl}$ for $X_j$. We mention that the approximation error here is also random because it depends on random quantities $N_j$ determined by thresholding.
Let $g_n$ denote the number of retained processes that may grow with the sample size $n$ in a nonparametric manner.
Recall that $V_j$ are the energies of processes. W.l.o.g., we assume for the moment that $V_1 \geq \dots \geq V_p$.
The following lemma quantifies $g_n$ and the number of retained coefficients $N_j$. 
One challenge is to deal with the discretization error with care when applying the concentration results.

\begin{lemma}\label{lemma:card}
Under Conditions \ref{cd:gauss}-\ref{cd:xsmooth}, \ref{cd:sampling}-\ref{cd:sn} and the weak $l_q$ sparsity, the number of retained processes $g_n \leq C \{m^{-1}\sqrt{\log p/n}\}^{-q/2}$ and the number of retained $\hat{\theta}_{ijl}$ for the $j$th process satisfies $N_j \leq C \{m^{-1}\sqrt{\log p/n}\}^{-1/(2\alpha+1)}j^{-2/\{q(2\alpha+1)\}}$ almost surely (a.s.) for some $C>0$.
\end{lemma}

\Cref{lemma:card} illustrates that many processes with small energies will be excluded from the estimation. The term $j^{-2/\{q(2\alpha+1)\}}$ indicates that the quantity $N_j$ will decrease as $V_j$ decays. Apparently, the processes will be screened out if $V_j$ decays to a smaller magnitude, i.e., $N_j$ will be zero for those processes. The retained coefficients of $X_j$ are thresholded from total $s_n$ terms, which to some extent implies a sufficiently large $s_n$.

\begin{theorem}[Approximation Error] \label{thm:apperror}
	Under the weak $l_q$ sparsity \eqref{bdecay}, if Conditions \ref{cd:egap}-\ref{cd:lqrn1} hold and $<\bpsi_k, \tilde{\bpsi}_{k} >_{\hilbert}$ $\geq 0$, then uniformly for $k=1,\dots,r_n$, we have the following.
	
	Case 1. When $q(2\alpha+1)>2$,
	\[
	\|\tilde{\bpsi}_{k} - \bpsi_k\|_{\hilbert} = O(k^{a+1}g_n^{1/2-1/q}), \qquad a.s.,
	\]
	
	Case 2. When $q(2\alpha+1)=2$,
	\[
	\|\tilde{\bpsi}_{k} - \bpsi_k\|_{\hilbert} = O\left[ k^{a+1} \{m^{-1}\sqrt{\log p/n}\}^{\alpha/(2\alpha+1)}(\log g_n)^{1/2}\right], \quad a.s.,
	\]
	
	Case 3. When $q(2\alpha+1)<2$,
	\[
	\|\tilde{\bpsi}_{k} - \bpsi_k\|_{\hilbert} = O\left[ k^{a+1} \{m^{-1}\sqrt{\log p/n}\}^{\alpha/(2\alpha+1)} \right], \qquad a.s..
	\]
	
\end{theorem}

\Cref{thm:apperror} establishes rates of convergence for approximation error based on the comparison of $\alpha$ and $q$ which represent sparsity levels within and between processes, respectively. 
The term $k^{a+1}$ is attributed to the increasing error of approximating higher order eigenelements $\bpsi_k$.
The approximation error is decomposed into two terms which incorporate errors caused by screening out processes with small energies and excluding coefficients with small variances for the retained processes.
Observe that smaller $q$ and larger $\alpha$ lead to sparser settings. When $\alpha$ is relatively large, saying $\alpha>1/q-1/2$ as in Case 1, the energies of processes $V_j$ do not decay so fast that the term $g_n^{1/2-1/q}$ caused by excluding the processes with small energies dominates. Intuitively in this case, the processes are more like scalar variables since the between-process sparsity dominates. When $q$ is relatively small, the rates are determined by the term $\{m^{-1}\sqrt{\log p/n}\}^{\alpha/(2\alpha+1)}$ attributed to thresholding coefficients of the retained processes, and the additional term $\log g_n$ in Case 2 is due to the fact that the $N_j$ corresponds to $j^{-2/\{q(2\alpha+1)\}}$ as a consequence of the decaying energies.

\begin{theorem}[Estimation Error] \label{thm:esterror}
	Under the weak $l_q$ sparsity \eqref{bdecay}, if Conditions \ref{cd:gauss}-\ref{cd:blip} and \ref{cd:lqrn2} hold and $<\hat{\bpsi}_k, \tilde{\bpsi}_{k} >_{\hilbert}$ $\geq 0$, then uniformly for $k=1,\dots,r_n$, we have the following.
	
	Case 1. When $\gamma > 1/(2-q)$,
	\[
	\|\tilde{\bpsi}_{k} - \hat{\bpsi}_k\|_{\hilbert} = O_p(k n^{-1/2}),
	\]
	
	Case 2. When $(1-\beta)/2 < \gamma \leq 1/(2-q)$ with $\log p/n = O(n^{\beta-1})$,
	\[
	\|\tilde{\bpsi}_{k} - \hat{\bpsi}_k\|_{\hilbert} = O_p(k^{a+1} g_n^{1/2} m^{-1}).
	\]
	
\end{theorem}

The estimation error does not involve the term $N_j$, as we quantify the discretization error of retained coefficients via retained processes using Bessel's inequality. 
The corresponding rate of convergence for the covariance of retained processes is of the order $O_p(n^{-1/2} + g_n^{1/2}m^{-1})$, where $g_n$ is the number of retained processes determined by quantities $q$ and $\gamma$ from  \Cref{lemma:card}.
Cases 1 and 2 correspond to the parametric covariance estimation error and discretization error, respectively. 
The rates of convergence exhibit a phase transition phenomenon depending on the sampling rate $\gamma$. 
When the data are sufficiently dense as in Case 1, the error term for covariance estimation induced by the discretization is negligible, achieving the parametric rate $n^{-1/2}$ as if the whole functions were observed. Using similar techniques in \citet{hall2007methodology}, we obtain a sharp bound for eigenfunctions. Otherwise as in Case 2, slower convergence rates for eigenfunctions by Theorem 1 in \citet{hall2006properties} are attained by taking the discretization error $m^{-1}$ into account.

Combining the approximation error and estimation error, one can see that the convergence rate of $\|\hat{\bpsi}_k - \bpsi_k\|_{\hilbert}$ can not exceed the parametric rate which is consistent with the common sense. The phase transition caused by smoothing has been discussed in \citet{cai2011optimal,cai2010nonparametric} and \citet{zhang2016sparse} for univariate functional data, while it is revealed for the first time for high-dimensional functional data. 


\section{Simulation Studies}\label{sec:sim}

\subsection{Sparse FPCA}

We conduct several experimental studies to illustrate the performance of the proposed method for high-dimensional functional variables. We first assess the estimators in an unsupervised fashion.

The noisy observations are generated from
$
y_{ij}(t_{k}) = x_{ij}(t_{k}) + \epsilon_{ijk} = \sum_{l=1}^{s} \theta_{ijl} \phi_l(t_{k}) + \epsilon_{ijk}, ~~t_{k} \in [0,1], j=1,\dots,p,
$
where $\epsilon_{ijk}$ are i.i.d. from $N(0, 1)$. Let $\phi_l(t)$ be functions in the Fourier basis, $\phi_l(t) = \sqrt{2} \sin\{\pi (l+1) t\}$ when $l$ is odd, $\phi_l(t) = \sqrt{2} \cos(\pi l t)$ when $l$ is even. We set $s=50$ to mimic the infinite nature of functional data. The equally spaced grids are $\{t_k\}_{k=1}^{m}=\{0, 0.01, \dots, 1\}$ with $m=101$, and the sample size $n =100$. Each simulation consists of 100 Monte Carlo runs.


To generate $x_{ij}(\cdot)$, define $w_{ij}(t) = \sum_{l=1}^s \tilde{\theta}_{ijl} \phi_l(t)$, where $\tilde{\theta}_{ijl} \stackrel{}{\sim} N(0, 16l^{-7/3})$ that are i.i.d across $i$ and $j$. The processes are given based on the autoregressive relationship,
\begin{equation*}
x_{ij}(t) = \sum_{j^{\prime}=1}^{p} \varrho^{|j-j^{\prime}|} j^{-1/q} w_{ij^{\prime}}(t) = \sum_{l=1}^{s}\sum_{j^{\prime}=1}^{p}\varrho^{|j-j^{\prime}|} j^{-1/q}\tilde{\theta}_{ij^{\prime}l} \phi_l(t) = \sum_{l=1}^{s} \theta_{ijl} \phi_l(t),
\end{equation*}
with $\theta_{ijl} = \sum_{j^{\prime}=1}^{p}\varrho^{|j-j^{\prime}|} j^{-1/q}\tilde{\theta}_{ij^{\prime}l}$. The constant $q$ determines the sparsity level and $\varrho$ controls the correlation among functional variables. Set $q=0.5$ and $\varrho = 0.5$. Let $p = 50, 100, 200$, respectively, for different experiments.

\begin{figure}[t!]
	\centering
	\subfigure[]{
		\begin{minipage}{0.3\linewidth}
			\includegraphics[width=1.8in]{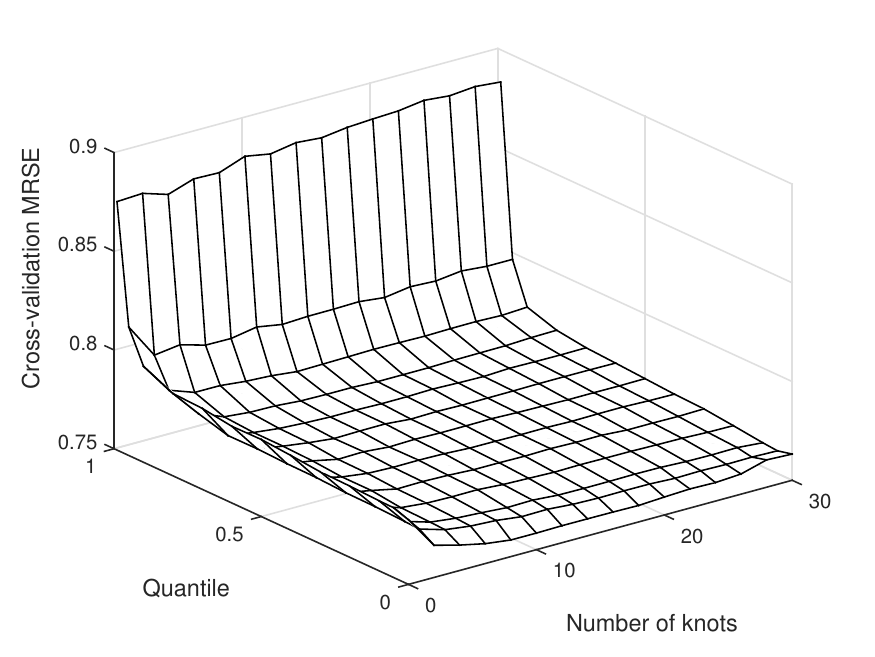}
		\end{minipage}
		\label{fig:simd}
	}
	\subfigure[]{
		\begin{minipage}{0.3\linewidth}
			\includegraphics[width=1.8in]{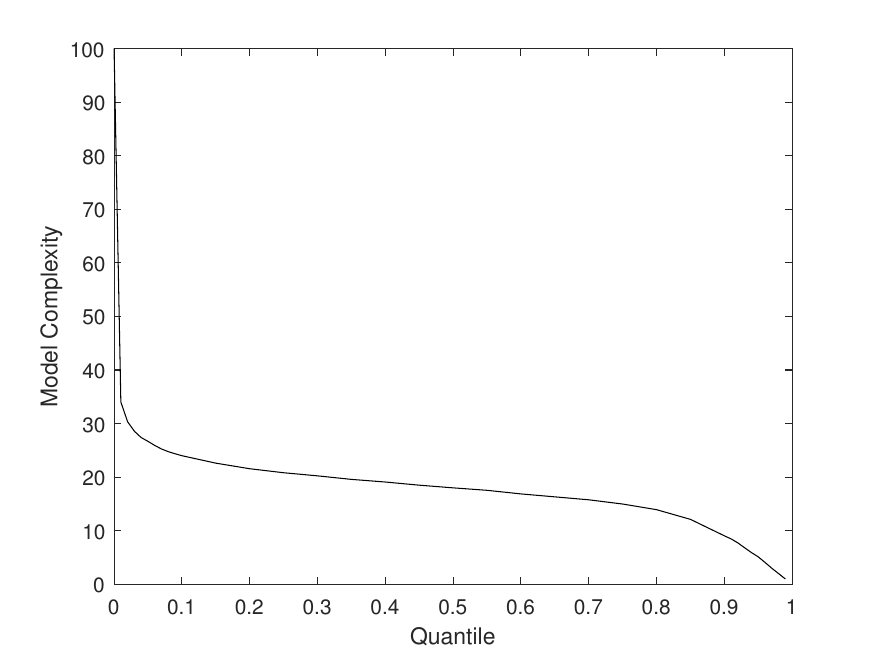}
		\end{minipage}
		\label{fig:sime}
	}
	\subfigure[]{
		\begin{minipage}{0.3\linewidth}
			\includegraphics[width=1.8in]{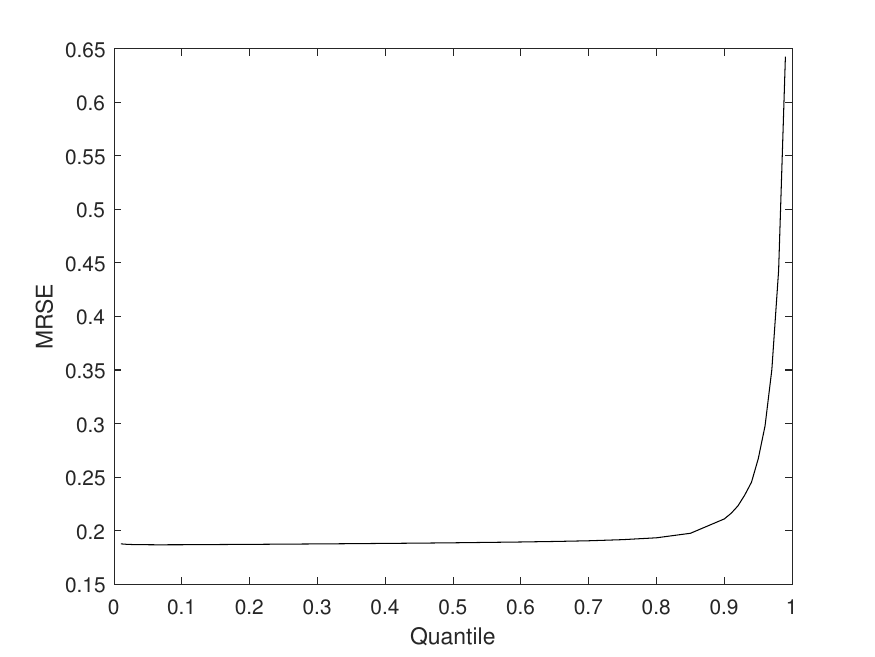}
		\end{minipage}
		\label{fig:simf}
	}
	\caption{The results for the weak $l_q$ sparsity setting with $p=100$: cross-validated MRSE under different quantile levels and different numbers of knots (a), model complexity (i.e., the number of retained processes) (b) and MRSE (c) under different quantile levels.  }
	\label{fig:mrse}
\end{figure}

To demonstrate the performance, we use the mean square error (MSE) for eigenfunctions $\|\bpsi(t) - \hat{\bpsi}(t)\|_{\hilbert}^2 = \sum_{j=1}^p \|\psi_j(t) - \hat{\psi}_j(t)\|^2$ and the mean relative square error (MRSE) for true curves, $n^{-1}\sum_{i=1}^{n} \| \bx_{i}(t) - \hat{\bx}_{i}(t) \|_{\hilbert}^2 / \|\bx_{i}(t)\|_{\hilbert}^2$. 
We use the number of retained processes to evaluate the model complexity. Moreover, we compare the results and computation time of our method to those of the HG method \citep{happ2018multivariate}.

We use orthonormal cubic spline basis for both methods. The results for $p=50$ revealing similar patterns are not presented for conciseness.
As for the parameters $s_n$ and $\rho$ in our method, it is computationally expensive to use cross-validation to choose both jointly.
Based on our experience, the results are actually not sensitive to $s_n$, as long as it is adequate, shown in \Cref{fig:simd}, but not too large for effective computation. This empirical finding is in line with our theory that it suffices to have an adequately large $s_n$. 
In particular, we use $s_n=14$ in the the $l_q$ setting for presented results.

\begin{table}[t!]
	\caption{The MSE with standard errors in parentheses for the first 4 eigenfunctions and the comparison of average computation time for a full sample recovery, where the quantile $\rho=0.5$ in our method.}
	\label{tab:sim-fpca}
	\centering
	\scriptsize
	\resizebox{\linewidth}{!}{
	\begin{tabular}{cccccc}
			\hline \hline
			\multicolumn{2}{c}{} & $\bpsi_1$ & $\bpsi_2$ & $\bpsi_3$ & $\bpsi_4$  \\ \hline
			\multirow{2}{*}{$p=100$} &sFPCA & .007(.005) & .031(.024) & .074(.046) & .242(.255) \\ 
			& MFPCA & .013(.005) & .059(.024) & .148(.047) & .381(.271) \\
			\multirow{2}{*}{$p=200$} & sFPCA & .007(.005) & .026(.016) & .073(.048) & .276(.254) \\ 
			& MFPCA & .019(.005) & .084(.019) & .211(.054) & .511(.320) \\ \hline
			\multicolumn{6}{c}{Average computation times for recovery (second)} \\ \hline
			& $s_n$ & 14 & 24 & 34 & 44  \\ 
			\multirow{2}{*}{$p=100$} & sFPCA & 1.269 & 2.099 & 3.210 & 4.464 \\ 
			& MFPCA & 7.366 & 26.52 & 68.68 & 139.4 \\ 
			\multirow{2}{*}{$p=200$} & sFPCA & 2.482 & 4.917 & 8.908 & 14.677 \\ 
			& MFPCA & 32.368 & 157.799 & 447.874 & 1017.838 \\ \hline
	\end{tabular} }
\end{table}

In such unsupervised problems, the influence of quantiles on the trade-off between the model complexity and quality of estimation/recovery is of main interest.
We obtain parsimonious models with satisfactory performance of recovery over a wide quantile range, see \Cref{fig:sime,fig:simf}.
As a practical advice, we suggest to choose a slightly large $\rho$ if model parsimony is of main concern. 
Briefly, in practice, we suggest first fix an adequately large $s_n$ and then determine the ``best'' choice of $\rho$. One might inspect performances of a few $s_n$ given the selected quantiles for confirmation. 

We see from \Cref{tab:sim-fpca} that our method with $\rho=0.5$ clearly outperforms the HG method, especially when $p$ is large. In comparison with sFPCA, the HG method includes all processes, which cannot yield parsimonious representations.
Lastly we illustrate substantial computational savings of our algorithm by reporting the average computation time over 100 Monte Carlo runs for a full sample recovery using different numbers of basis functions on a standard computer with 2.40GHz I7 Intel microprocessor and 16GB of memory, see \Cref{tab:sim-fpca}. The results roughly agree with the computation complexity  $O(nps_n+nN^2+N^3)$ for our approach and $O(np^2s_n^2+p_n^3s_n^3)$ for the HG method in \Cref{rem:complexity}, where $N=\sum_{j=1}^p N_j$ quantifies the number of all retained coefficients after thresholding that often entails $N\ll ps_n$.

\begin{table}[t!]
	\caption{The averages of misclassification rates on testing samples with standard errors in parentheses across different $r_n$ and the average computation time. Also in square brackets shown are the average model complexity of the proposed method with standard errors in parentheses.}
		\label{tab:classify}
		\centering
		\resizebox{\linewidth}{!}{
		\begin{tabular}{ccccccc}
			\hline \hline
			\multirow{2}{*}{Method} & \multicolumn{5}{c}{$r_n$} & \multirow{2}{*}{Time (second)} \\ \cline{2-6}
			& 2 & 5 & 8 & 12 & 15 & \\ \hline  
			\multirow{2}{*}{\begin{tabular}[c]{@{}c@{}}sFPCA\\ +LDA\end{tabular}} & \multirow{2}{*}{\begin{tabular}[c]{@{}c@{}}30.19(3.78) \\ {[}2.62(4.88){]}\end{tabular}} & \multirow{2}{*}{\begin{tabular}[c]{@{}c@{}}13.41(2.79) \\ {[}2.47(5.59){]}\end{tabular}} & \multirow{2}{*}{\begin{tabular}[c]{@{}c@{}}13.14(2.68) \\ {[}2.49(5.41){]}\end{tabular}} & \multirow{2}{*}{\begin{tabular}[c]{@{}c@{}}13.66(2.78) \\ {[}2.54(6.26){]}\end{tabular}} & \multirow{2}{*}{\begin{tabular}[c]{@{}c@{}}14.09(2.82) \\ {[}2.62(6.48){]}\end{tabular}} & \multirow{2}{*}{1.28} \\
			&  &  &  &  & & \\ 
			\multirow{2}{*}{\begin{tabular}[c]{@{}c@{}}MFPCA\\ +LDA \end{tabular}} & \multirow{2}{*}{30.66(3.83)} & \multirow{2}{*}{15.55(2.77)} & \multirow{2}{*}{14.75(2.74)} & \multirow{2}{*}{14.67(2.79)} & \multirow{2}{*}{14.68(2.59)} & \multirow{2}{*}{7.78} \\
			&  &  &  &  &  & \\ 
			\multirow{2}{*}{\begin{tabular}[c]{@{}c@{}}UFPCA\\ +ROAD \end{tabular}} & \multirow{2}{*}{34.27(5.77)} & \multirow{2}{*}{17.53(8.31)} & \multirow{2}{*}{16.46(8.04)} & \multirow{2}{*}{16.53(7.83)} & \multirow{2}{*}{16.55(7.94)} & \multirow{2}{*}{42.05} \\
			&  &  &  &  & &  \\ \hline
	\end{tabular}}
\end{table}

\subsection{Classification}

We inspect the performance of our algorithm on subsequent classification. The data are generated from $y_{ij}^{(\ell)}(t_{ijk}) = \mu_j^{(\ell)}(t) + x_{ij}^{(\ell)}(t_{ijk}) + \epsilon_{ijk}$ where $\ell=1$ or 0 denotes class 1 or 0, respectively. Let $\kappa$ denote the number of significant processes for classification. We set $\mu_j^{(0)}(t) =0$ for $j=1,\dots,p$ and $\mu_j^{(1)}(t)$ are linear combinations of the first 5 eigen-functions with weights equal to $(1, 1, -0.75, 0.75, 0.5)$ for $j=1,\dots,\kappa$, and the rest $\mu_j^{(1)}(t)=0$ for $j=\kappa+1,\dots,p$. We set $\kappa=2$ and $p=100$. The coefficients $\{\theta_{ijl}^{(\ell)}\}$ for both groups follow the previous generation mechanisms with slight modification: $\tilde{\theta}_{jl}^{(\ell)} \sim N(0, 3 l^{-2}), j=1,\dots,p, l=1,\dots,s$.
In each of 100 Monte Carlo runs, we generate a training set of 100 subjects and an independent testing set of 200 subjects, where half of these belong to each class. The proposed method and HG method both obtain $r_n$ multivariate scores $\hat{\eta}_{ik} = \sum_{j=1}^p \int_0^1 y_{ij}^{*}(t) \hat{\psi}_{kj}(t) dt$ which are low-dimensional and allow to apply the classical linear discriminant analysis (LDA) for classification. We also consider another viable method which combines and trains the scores obtained from univariate FPCA for $p$ processes with the high-dimensional classifier ROAD proposed by \citet{fan2012road}. 

In the supervised problem, we tune $s_n$ and $\rho$ jointly by 5-fold cross-validation, and choose the parameters of other methods in a similar manner. For comprehensive comparison, we train the models by retaining $2, 5, 8, 12, 15$ principal components, respectively. 
The principal components mean multivariate scores $\eta_{k}$ for the first two methods and univariate scores $\xi_{jk}$ for the last one.
As shown in \Cref{tab:classify}, the parsimonious models obtained by our method enjoy favorable classification performance. 
Our algorithm successfully selects relevant processes in nearly all runs, while the HG method treats all processes equally and fails to distinguish important processes. Although the last method adopts a high-dimensional classifier, it still performs worse than our approach. Furthermore, the average computation time over different $r_n$ and 100 Monte Carlo runs is reported, where chosen parameters are used for our approach and the HG method, and the R package `fdapace' is used for implementing the univariate FPCA. The result indicates that our proposal is much more 
computationally efficient for high-dimensional functional data.

\section{Real Data Example}\label{sec:realdata}

\begin{figure}[t!]
	\label{fig:EEG}
	\centering
	\subfigure[]{
		\begin{minipage}{0.45\linewidth}
			\includegraphics[width=2in]{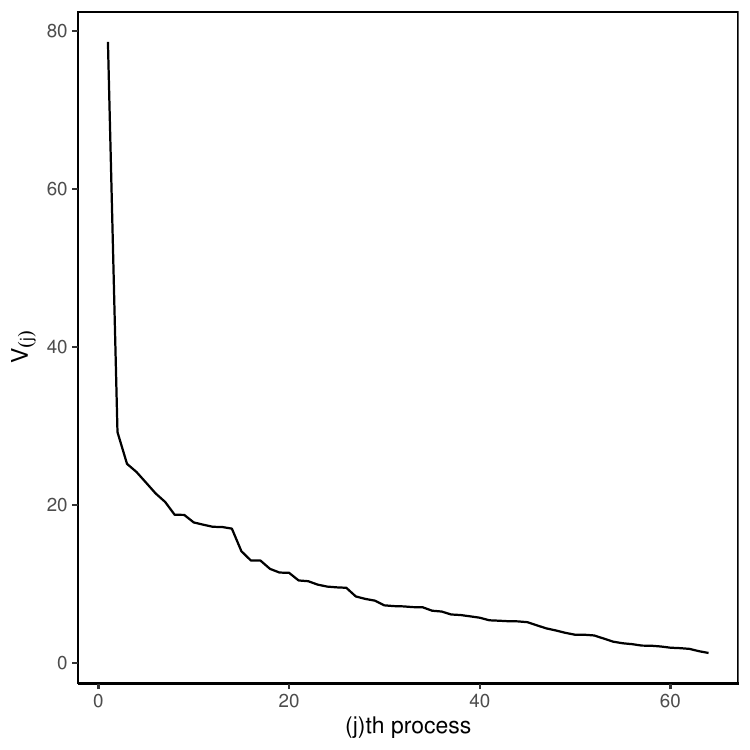}
		\end{minipage}
		\label{fig:eega}
	}
	\subfigure[]{
		\begin{minipage}{0.45\linewidth}
			\includegraphics[width=2.25in]{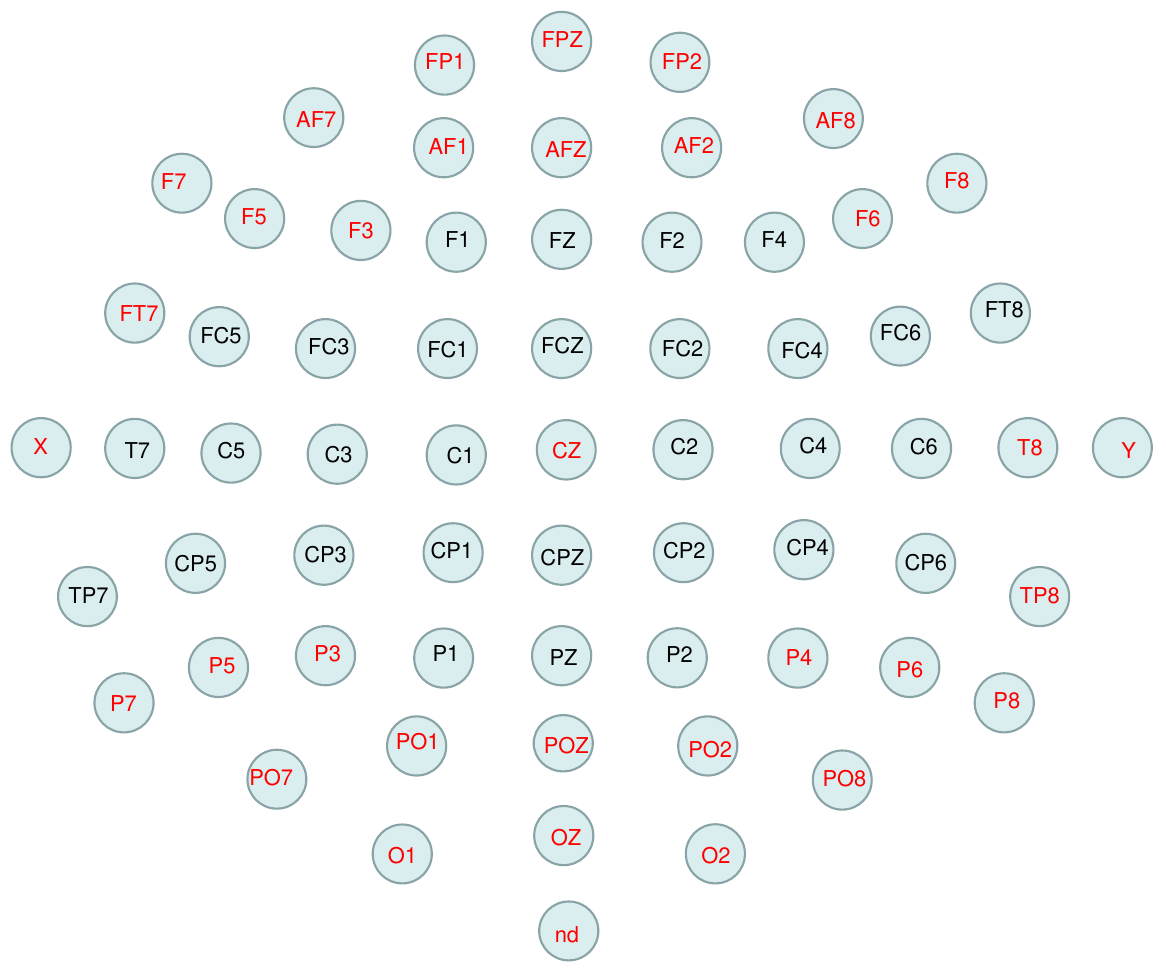}
		\end{minipage}
		\label{fig:eegb}
	} 
	\caption{(a) the ordered energies $V_{(j)}$ of EEG data. (b) the electrode names and positions where the ones marked in red are selected by our method with chosen parameters over a half runs.}
\end{figure}

We apply the proposed method to the electroencephalography (EEG) data obtained from an alcoholism study \citep{zhang1995event,ingber1997statistical}.  The data consists of $n=122$ subjects, 77 in the alcoholic group and 45 in the control group with each exposed to either a single stimulus or two stimuli. There are 64 electrodes placed at standard locations on the participants' scalp to record the brain activities. Each electrode is sampled at 256 HZ for one second interval. Hence each subject involves $p=64$ different functions observed at 256 time points. 
This dataset contains high-dimensional functional processes and was analyzed for functional graphical models \citep{qiao2019functional,qiao2020doubly,solea2020copula}. \citet{hayden2006patterns} found evidence of regional asymmetric patterns between the two groups by using 4 representative electrodes from the frontal and parietal regions.


\begin{table}[t!]
	\caption{The average misclassification rates on testing samples and computation time with standard errors in parentheses across different number of eigenfunctions. Also in square brackets shown are the average model complexity of sFPCA with standard errors in parentheses.}
	\label{tab:eeg}
	\centering
	\resizebox{\linewidth}{!}{
		\begin{tabular}{ccccccc}
			\hline \hline
			\multirow{2}{*}{Method} & \multicolumn{5}{c}{$r_n$} & \multirow{2}{*}{Time (second)} \\ \cline{2-6}
			& 10 & 20 & 30 & 40 & 50 & \\  \hline
			\multirow{2}{*}{\begin{tabular}[c]{@{}c@{}}sFPCA\\ +LDA\end{tabular}} & \multirow{2}{*}{\begin{tabular}[c]{@{}c@{}} 14.25(3.98) \\ {[}34.08(17.34){]}\end{tabular}} & \multirow{2}{*}{\begin{tabular}[c]{@{}c@{}} 14.73(3.46) \\ {[}36.07(19.47){]}\end{tabular}} & \multirow{2}{*}{\begin{tabular}[c]{@{}c@{}} 13.68(3.54) \\ {[}37.12(16.77){]}\end{tabular}} & \multirow{2}{*}{\begin{tabular}[c]{@{}c@{}} 13.18(3.87) \\ {[}35.19(16.64){]}\end{tabular}} & \multirow{2}{*}{\begin{tabular}[c]{@{}c@{}} 13.28(3.55) \\ {[}33.30(16.10){]}\end{tabular}} & \multirow{2}{*}{0.31} \\
			&  &  &  &  & &  \\ 
			\multirow{2}{*}{\begin{tabular}[c]{@{}c@{}}MFPCA\\ +LDA\end{tabular}} & \multirow{2}{*}{19.38(4.53)} & \multirow{2}{*}{19.05(4.33)} & \multirow{2}{*}{18.40(4.21)} & \multirow{2}{*}{17.05(4.54)} & \multirow{2}{*}{17.33(4.34)} & \multirow{2}{*}{3.74} \\
			&  &  &  &  & & \\
			\multirow{2}{*}{\begin{tabular}[c]{@{}c@{}}UFPCA\\ +ROAD\end{tabular}} & \multirow{2}{*}{16.50(4.10)} & \multirow{2}{*}{16.05(4.19)} & \multirow{2}{*}{16.10(4.21)} & \multirow{2}{*}{16.10(4.21)} & \multirow{2}{*}{16.10(4.21)} & \multirow{2}{*}{364.18} \\
			&  &  &  &  &  & \\ \hline
			TSROAD & \multicolumn{5}{c}{34.30 (0.06)}  & 138.85 \\ \hline
	\end{tabular}}
\end{table} 

We consider the average recordings for each subject under the single stimulus condition.
As shown in Figure \ref{fig:eega}, the energies $V_{(j)}$ exhibit a sparsity pattern, which indicates that the sparsity assumption is advisable in practice for high-dimensional functional data.
Our goal is to classify alcoholic and control groups based on their recordings. 
For each group, we randomly select two thirds of participants as the training sample and the rest as the test sample. We repeat 100 times and use the three methods in simulation, as well as  {the tangent space linear discriminant analysis method \citep{barachant2011multiclass} coupled with ROAD (termed as TSROAD) as the dimension of the tangent space $p(p+1)/2$ is large, to evaluate the classification performance.}
Due to sample splitting, the sample size of training samples is rather small, especially for the control group. Thus we calculate the misclassification errors over a candidate set of parameters in each method and use the lowest for comparison. Table \ref{tab:eeg} presents the misclassification rates for all considered methods under several $r_n$, indicating the superiority of our method with minimal misclassification errors. In particular, the TSROAD performs poorly, indicating substantial discriminative information loss which might be due to averaging over time. 
Moreover, the average computation time in Table \ref{tab:eeg} demonstrates the scalability of our approach for large $p$ and $m$, which is consistent with the computation complexity discussed in Remark \ref{rem:complexity}.
The Figure \ref{fig:eegb} presents the 64 electrode names and positions, and the electrodes marked in red indicate the ones selected more than half of 100 runs by our method with chosen parameters. It is observed that the retained electrodes mainly lie in the frontal and parietal regions. 

\section*{Acknowledgement}
Fang Yao's research is supported by the National Natural Science Foundation of China Grants 11931001 and 11871080, the LMAM and the Key Laboratory of Mathematical Economics and Quantitative Finance (Peking University), Ministry of Education.

\fontsize{9}{14pt plus.8pt minus .6pt}\selectfont

\bibliographystyle{asa}
\bibliography{sFPCA}

\end{document}